\documentclass[useAMS,usenatbib]{mn2emdf}

\usepackage{graphicx}

\def \mref#1{(\ref{#1})} 
\newcommand\sss{\scriptscriptstyle}

\def \rhomu {\rho_{\rm rt}}
\def \rhob {\rho_B}
\newcommand{\rlc}{R_{\rm lc}}
\newcommand{\nobs}{\hat n_{\rm obs}}
\newcommand\phf{\phi_f}
\newcommand\phpa{\phi_{\rm PA}}
\newcommand\phprof{\phi_{\rm prof}}
\newcommand\phobs{\phi_{\rm obs}}
\newcommand\phobsi{\phi_{{\rm obs,}i}}
\newcommand\phz{\phi_0}
\newcommand\phemab{\phi_{\rm em}}
\newcommand\phem{\phi^\prime_{\rm em}}
\newcommand\psir{\psi_r}
\newcommand\psira{\psi_{\sss \Omega}}
\newcommand{\thlo}{\theta_{\rm lo}}
\newcommand{\thzc}{\theta_{\rm zc}}
  \newcommand{\rn}{r/\rlc}
  \newcommand\om{(\vec \Omega, \vec \mu)}
  \newcommand\omobs{(\vec \Omega, \nobs)}
  \newcommand\thpc{\theta_{\rm pc}}
  \newcommand\dphbcw{\Delta\phi_{\rm BCW}}
  \newcommand\phmu{\phi_{\mu}}
  \newcommand\rmin{r_{\rm min}}
  \newcommand\rmax{r_{\rm max}}
\def \rns{R_{\rm NS}}

\title[Altitude-dependent polarization in radio pulsars]
{Altitude-dependent polarization in radio pulsars}

\author[J. Dyks]{J. Dyks\\
Nicolaus Copernicus Astronomical Center, Toru\'n, Poland}
\begin{document}

\date{Accepted 1988 December 15. Received 1988 December 14; in original form 1988 October 11}

\pagerange{\pageref{firstpage}--\pageref{lastpage}} \pubyear{2002}

\maketitle

\label{firstpage}

\begin{abstract}
Because of the corotation, the polarization angle (PA) curve of a pulsar
lags the intensity profile by $4r/\rlc$ rad in pulse phase.
I present a simple and short derivation of 
this delay-radius relation to show that it is not caused by the aberration
(understood as the normal beaming effect) but purely by contribution of
corotation to the electron acceleration in the observer's frame.
Available altitude-dependent formulae for the PA curve
are expressed through observables and emission altitude 
to make them immediately ready to use in radio data modelling.
The analytical approximations for the altitude-dependent PA curve
are compared with exact numerical results to show how they perform
at large emission altitudes. 
I also discuss several possible explanations for the opposite-than-normal
shift of PA curve, exhibited by the pedestal emission of
B1929$+$10 and B0950$+$08.
\end{abstract}

\begin{keywords}
pulsars: general -- pulsars: individual: B1929+10 --
B0950+08 -- Radiation mechanisms: non-thermal.
\end{keywords}

\section{Introduction}

In the simplest model of pulsar polarization
(Komesaroff 1970; Radhakrishnan \& Cooke 1969, hereafter RC69)
the position angle of polarization does not depend on the radial 
distance\footnote{The quantity $r$ represents the distance
measured from the \emph{center} of the neutron star.}
of radio emission $r$.
The polarization angle (PA) 
becomes dependent on $r$ when dynamic effects
of pulsar's rotation are taken into account (dragging of electrons
by the corotating magnetic field).
Blaskiewicz et al.~(1991) (hereafter BCW)
showed that if the emission originates
from a fixed radial distance $r$, the shape of the PA swing
is (approximately) preserved (ie.~it is the same as in the case of 
negligible $r$),
but the entire swing is shifted towards later phases by
$\dphbcw\approx4r/\rlc$ radians with respect to the
center of the profile (where $\rlc$ is the light cylinder radius).
Following BCW, I will refer to this formula 
with the term `delay-radius'
relation.
Hibschmann \& Arons (2001) (hereafer HA)
have shown that the PA curve also undergoes
vertical shifts, ie.~in the PA values.
Both these results appear to have interesting observational
consequences (eg.~Ramachandran \& Kramer 2003; von
Hoensbroech \& Xilouris 1997).

In the case of phase-dependent emission altitude different parts of 
the PA curve undergo different shifts and the PA curve assumes 
a distorted shape. This effect regularly happens to be employed
to model observed distortions of PA curves and to derive
magnetospheric emission altitudes
(eg.~Krishnamohan \& Downs 1983, hereafter KD83;
Xu, Qiao \& Han 1997; Gil \& Krawczyk 1997; Mitra \& Seiradakis 2004).
A tool that is needed for this is an analytical formula
for the PA that explicitly depends on the radial distance of the emission
region. BCW and HA provide various forms of this equation. However,
their formulae are not in a ready-to-use form:
they are expressed through the emission time instead of the pulse
longitude $\phobs$ (hereafter called pulse 
\emph{phase}\footnote{Throughout this paper the phase is assumed 
to be measured in radians
whenever dimensionless terms are added to it.}).
It is the need for this last step of the BCW's analysis 
that actually sparked writing of this paper.

A strict and formal description of the PA subject
is given in the superb work of Hibschman \& Arons 
(2001; see their appendices) and it will not be repeated in this paper. 
My intention here is to provide a simple reference for those who
want to use the altitude-dependent PA curves in their data modelling.
The aim is to clarify some obscure aspects
of the subject by trivialising the formalism and to provide practical 
PA equations
in their final form.
Accordingly,
Sect.~\ref{simpderiv} presents a very simple
and short derivation
of the delay-radius relation to clearly expose its origin.
In Sect.~\ref{pacurv} I introduce the fiducial phase, describe
the magnitudes and directions of various relativistic shifts
with respect to it, and
I write down 
the equations for the altitude-dependent PA curve in a form that is ready
for immediate use in data modelling. 
In Sect.~\ref{limits} I compare the analytical approximations to exact
numerical results obtained for various emission altitudes to show the validity range
of the BCW theory. In Sects.~\ref{accuracy} and \ref{noncurv}
I discuss possible explanations for the opposite-than-expected shifts
of PA curve, as exemplified by the pedestal radio emission components
of B1929$+$10 and B0950$+$08. These and other interpretations of the anti-BCW 
shifts are summarized in  Sect.~\ref{antibcw}.

\section{Simple derivation of the delay-radius relation
due to corotation}
\label{simpderiv}

The derivation presented here is constrained  
to the case of the equatorial\footnote{Unless
specified otherwise, `equatorial' refers to the \emph{rotational} equator.}
plane of an orthogonally rotating pulsar 
(with dipole inclination $\alpha=90^\circ$).
However, it is simple, intuitive and demonstrates the effect more
directly than the original derivation of BCW.

Because of their corotation, trajectories of electrons
are bent forward (toward the direction of rotation) in the inertial 
observer frame (IOF) with respect to trajectories in the corotating frame (CF).
The trajectory of electrons that move along
the dipole axis acquires some forward
curvature, whereas the rectilinear motion occurs somewhere
on the trailing side of the dipole axis. The bundle of IOF-trajectories
becomes approximately symmetric around the location of this straight trajectory,
which now assumes the role that the dipole axis had in the 
RC69 model (RVM model). 
This zero-curvature trajectory is associated 
with the inflection point of the PA curve.
Its location can be found in the following two steps. First, we
calculate `rotationally-induced curvature', which is understood as
the curvature imposed by rotation on a trajectory that 
is a straight line in the CF. Second, we will search for
a place on the trailing side of the dipole axis, where this rotational
curvature is exactly cancelled by 
the same-magnitude backward curvature of dipolar magnetic 
field lines.

Relativistic electrons 
(with the speed of essentially constant magnitude $v \simeq c$)
that move radially along the dipole axis 
in the frame rotating with the neutron star's angular velocity $\vec \Omega$,
undergo acceleration $a \simeq 2\Omega c$ in the IOF.
This result can be obtained in several elementary ways
(see the exercise in Appendix \ref{ladyb}). 
The particle trajectory along the dipole axis thus acquires radius 
of curvature
$\rhomu = c^2/a$ which is:
\begin{equation}
\rhomu \simeq \frac{\rlc}{2}.
\label{rheq}
\end{equation}
The index `rt' is to remind that this is the curvature induced by rotation
on electron \emph{trajectory} that is \emph{radial} (or almost radial) 
in the corotating frame.
Near the dipole axis, the curvature radius of dipolar field lines is given by
\begin{equation}
\rhob \simeq \frac{4}{3}\frac{r}{\sin\theta},
\end{equation}
where $r=|\vec r|$ is the radial distance of an emission point
and $\theta$ is the angle between $\vec r$ and the dipole axis.

We look for regions in the trailing part of the polar cap tube, where
the rotationally-induced curvature of eq.~\mref{rheq}
is cancelled by the curvature of
B-field lines. By equating $\rhomu$ with $\rhob$ one can find 
an equation for the locations of straight portions of 
the trajectories:
\begin{equation}
\sin\thzc \simeq \thzc\simeq \frac{8}{3}\frac{r}{\rlc},
\label{zerocurv}
\end{equation}
where the index `zc' stands for `zero curvature'. 
We can see that the 
locations of the zero-curvature depend on radial distance: $\thzc \propto r$.

In the dipole geometry, magnetic field lines at points with coordinate $\theta$
are directed at angle $\theta_k \simeq (3/2)\theta$ with respect to the
magnetic axis.
Therefore, the phase delay of emission
from the zero-curvature regions with respect to the (same altitude) 
emission along dipole axis is equal to 
\begin{equation}
\dphbcw \simeq \theta_k\left(\thzc\right) \simeq \frac{3}{2}\ \thzc \simeq 4\frac{r}{\rlc}.
\label{bcw}
\end{equation}
Thus, the locally-straight portions of electron trajectory
lag the dipole axis spatially by $\thzc\simeq (8/3)r/\rlc$, whereas the
tangent-to-$B$ directions of radio waves emitted from these regions lag 
the direction of waves emitted along the dipole axis by the angle of 
$4r/\rlc$ radians.
Note that until the very end of the derivation 
there was no need to explicitly refer to the
effects of aberration and propagation time delays (APT) that
noticeably increase complexity of more general analysis
(see Appendix \ref{abret} for the definition
of the APT effects).
This is because it is the rotational straightening 
of electron trajectories that is the essence of the effect.
As long as one is interested in emission from a fixed altitude,
\emph{both} the aberration \emph{and} propagation time effects 
can be ignored,
because the waves emitted along the dipole axis as well as those emitted 
from the zero-curvature regions are advanced in (absolute) phase
by roughly the same magnitude ($r/\rlc$ by aberration and another $r/\rlc$
by propagation time). It is so because in the small angle approximation
all the open field lines are basically orthogonal 
to the corotation velocity (for details see Dyks, Rudak \& Harding 2004b,
hereafter DRH).
The APT effects only have to be included when there are altitude differences,
or when one wants to know the magnitude of absolute delays with respect to the
``fiducial phase". This subject will be discussed in detail in the next section.

According to eq.~(\ref{zerocurv}), the last open field lines, 
located near $\thlo \simeq (r/\rlc)^{1/2}$,
have the zero-curvature in IOF at radial distance 
\begin{equation}
\frac{r}{\rlc} \simeq \frac{9}{64} \simeq 0.14. 
\end{equation}
Note, however, that at this altitude the inaccuracy 
of our approximation (linear in $r/\rlc$)
is \emph{not} negligible (it is of the order of $(r/\rlc)^{1/2} \sim 40\%$).

We can also determine the rotation period $P=2\pi/\Omega$
for which the zero curvature occurs at the trailing part of the rim 
of the polar cap.
Since $\thzc/\thlo \simeq (8/3)(r/\rlc)^{1/2} = 0.0386 (r_6/P)^{1/2}$,
with $r_6 =r/(10^6{\rm\ cm})$ and $P$ in seconds,
one obtains:
\begin{equation}
P \simeq 1.49 \cdot 10^{-3}\ {\rm s}\  R_6,
\end{equation}
where $R_6 = \rns/(10^6{\rm\ cm})$ is the neutron star radius.
Thus, it happens just in the fastest known millisecond pulsars
(J17148$-$2446 with $P=1.396$ ms, Hessels et al.~2006, 
or B1937$+$21 with $P=1.558$ ms, Backer et al.~1982)
that the `dipole axis' is shifted full way to the trailing rim of the polar cap
(provided the pulsars have $\alpha \sim 90^\circ$, as their interpulses 
suggest). 
The effect is illustrated in Fig.~1 of Dyks \& Rudak (2002),
where a numerically calculated trajectory of photons in the CF
(for $P=1.5$ ms and $\alpha=90^\circ$)
neatly coincides with the last open field line that emerges from
the trailing side of the polar cap. 
Another consequence is that the efficiency of magnetic pair production
in these objects is the weakest near the trailing side of the polar cap,
not at the polar cap center (see Fig.~7 in the last-mentioned paper).

\section{Polarization angle curve}
\label{pacurv}

In the absence of the rotational effects the PA as a function of phase
is given by:
\begin{equation}
\psi = \tan^{-1}\left[\frac{\sin\alpha\sin(\phobs-\phz)}
{\cos(\phobs-\phz)\cos\zeta\sin\alpha-\cos\alpha\sin\zeta}\right] + \psira,
\label{orig}
\end{equation}
where $\phobs$ is the pulse phase with zero point defined arbitrarily
by the observer, $\phz$ is the pulse phase at which the line of sight
lies within the $\om$ plane, the constant $\psira$ is the
position angle of the projection of the pulsar rotation axis
on the plane of the sky, $\alpha$ is the angle between the angular
velocity of pulsar rotation $\vec \Omega$ and the dipole 
magnetic moment $\vec \mu$,
whereas $\zeta$ is the angle between $\vec \Omega$ and the unit vector
of the line of sight $\nobs$.
The sign of the $\arctan$ term is for $\psi$ measured 
in the observers' conventional way, i.e~\emph{counterclockwise}
in the sky (Everett \& Weisberg 2001).
In this simple geometrical model, the PA is determined purely by projection of
magnetic field direction on the sky's plane. Therefore, the azimuth 
angle $\phmu$ of the dipole axis
(measured from the plane which contains an observer and $\vec \Omega$, 
see Fig.\ref{emigeom})
and the pulse phase $\phobs$ are simply related by $\phmu = \phobs - \phz$.
Let us denote the value of pulse phase 
at which the center
of the PA curve (its inflection point) is observed by $\phpa$
(at this phase $|d\psi/d\phobs|$ has a maximum and $\psi=\psira$).
In the absence of the rotational effects,
(and in the case of infinite propagation speed $c =\infty$),
the center of the PA curve would be observed at the pulse phase $\phpa = \phz$.
Let us use the symbol $\phprof$ to denote the center of the pulse profile
defined as the midpoint between the profile's outer edges.
As long as the outer boundary of the radio emission region in the CF
is symmetric with respect to the $\om$ plane, the center of
the pulse profile would be observed at exactly the same phase
$\phprof=\phz$.

\subsection{Altitude-dependent polarization angle curve}

   \begin{figure}
      \centering
      \includegraphics[width=0.5\textwidth]{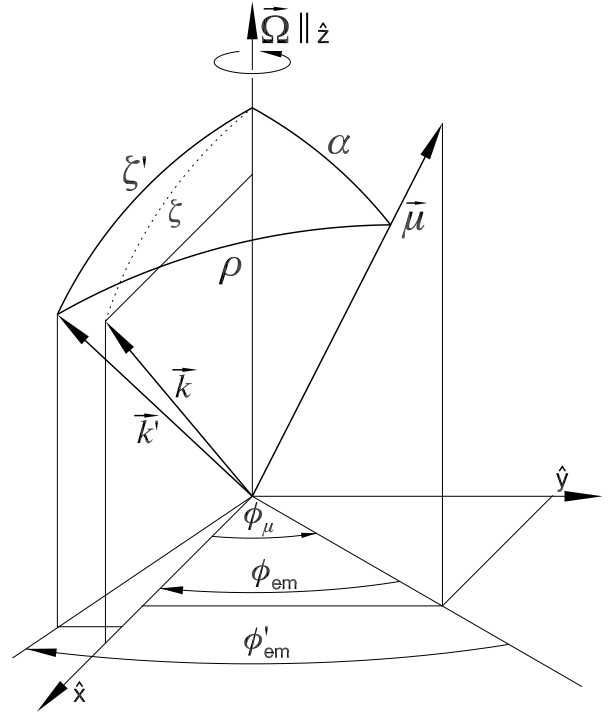}
      \caption{Orientation of the CF emission direction $\vec k^\prime$,
      IOF emission direction $\vec k$, and the dipole axis $\vec \mu$
      at the moment when radiation emitted at angle $\rho$ with respect 
      to $\vec \mu$ in the CF becomes directed toward an observer who is
      located
      at angle $\zeta$ from the rotation axis in the $(\vec \Omega, \hat x)$
      plane. The line of sight direction $\nobs$ 
      coincides with $\vec k$.
      Up to the order of $r/\rlc$ the angles $\zeta$ and $\zeta^\prime$ 
      can be considered equal. The azimuths $\phemab$ and $\phem$
      differ by $r/\rlc$.
      Note that the angles $\phemab$ and $\phem$ are assumed to increase
      in the opposite direction than $\phmu$.
      }
      \label{emigeom}
   \end{figure}

BCW generalized eq.~(\ref{orig}) to include rotational effects.
By considering the radio emission \emph{from a fixed
 radial distance $r$}, they found that
 the PA curve is shifted toward later
 phases by $\dphbcw\approx 4\rn$ \emph{with respect to the center of the pulse
 profile}. For this special case of fixed $r$, the profile's center 
 is a very convenient reference point from the
 observational point of view: if measured, the shift can be directly
 translated into the radial distance $\rn \approx \dphbcw/4$.\footnote{To find
 the phase of inflection point $\phpa$ one can fit the standard PA swing
(eq.~\ref{orig}) to the data (as BCW and von Hoensbroech \& Xilouris did),
because it has nearly the same shape as the equation which includes
the rotational effects.}
For other applications of BCW theory, however, the center of the pulse
profile is not a suitable reference point,
because the phase at which it occurs ($\phprof$) depends on $r$
itself. For example, the shift (PA center - profile center) does not tell us
by how much the PA curve is shifted with respect to the case with
the rotational effects ignored. Neither is the profile center helpful
if different parts of the profile originate from different altitudes.

A good reference point for measuring
altitude-dependent shifts in pulse profiles should be unambiguously associated
with the azimuth $\phmu$ of the magnetic dipole axis.
This criterion is met by the fiducial phase $\phf$, defined as follows:
it is the pulse phase at which the observer detects a photon that was
emitted
from the center of the star at the moment when the dipole axis was 
in the plane containing
$\vec \Omega$ and the observer (hereafter $\omobs$-plane). 
Because $\phobs = \phf$ corresponds to
$\phmu = 0$ \emph{strictly in the just-described sense}, 
the fiducial phase $\phf$
plays a similar role as $\phz$ does
in eq.~\ref{orig} (ie.~one may consider them identical: $\phf \equiv \phz$).

The shape of emission region that I consider in this paper
is assumed to be axially symmetric around the dipole axis,
ie.~its radial distance $r$ can be described
as a function of only the angle $\rho \approx 1.5s(r/\rlc)^{1/2}$
between the emission direction in the CF and the dipole axis
($s=\theta_{\rm surf}/\thpc$ is the footprint parameter
of a $\vec B$-field line on the star surface).
Such an \emph{average} form is roughly consistent with 
the observed shapes of phase-averaged pulse profiles 
(eg.~Johnston et al.~2008;
Rankin 1983) but neglects
the azimuthal structure (Karastergiou \& Johnston 2007;
Rankin \& Ramachandran 2003).
To calculate the PA curve for an arbitrary shape of $r(\rho)$ 
one needs an equation for the
altitude-dependent PA curve. Equations (16) or (17), (the latter with $\phz =
3\rn$), from BCW can be used for this purpose, provided that a
transition from `$\Omega t$' to the pulse phase $\phobs$ is carefully done.

The equations (16) and (17) of BCW are expressed in the emission 
time\footnote{The emission time refers to a Lorentz frame in which 
the pulsar's center
of mass is at rest. Note that BCW use two different symbols to denote it
($t$ as well as $t_e$).} \emph{with the zero point defined in a particular
way:} $t\equiv t_e=0$
corresponds to the moment when the dipole axis is in the $\omobs$-plane. 
Thus, the emission time in BCW is simply defined as $t_e = \phmu/\Omega$,
where $\phmu$ is the azimuth of $\vec \mu$ in the frame with $\vec \Omega
\parallel \hat z$ and with the observer in the $(\vec \Omega, \hat x)$ plane
(see Fig.~\ref{emigeom}; this azimuth provides the measure of the emission 
time).
It is very important to discern the emission time (or $\phmu$) from
the detection time $t_d$ (or from pulse phase $\phobs$) at which the radiation
emitted at $t_e$ is detected. They are related by: 
\begin{equation}
t_d = t_e + d/c - \vec r \cdot \nobs / c
+ \Delta t_{\rm zp},
\end{equation}
where $\nobs$ is the unit vector pointing towards the observer, $d$ is
the pulsar's distance and $\Delta t_{\rm zp}$ takes into account the fact that
the observer is allowed to assume arbitrary zero point in the counting of time.
The term $\vec r \cdot \nobs/c$ (`propagation time advance') 
takes into account the fact that the source
located at $\vec r$ is closer to the observer than the center of the
neutron star.
The same can be expressed in terms of angles:
\begin{equation}
\phobs = \phmu + \Omega d/c -
\Omega \vec r \cdot \nobs / c + \Delta\phi_{\rm zp},
\label{phobs1}
\end{equation}
 where $\phobs = \Omega t_d$, and $\Delta\phi_{\rm zp}$
 takes into account the fact that the observer can assign phase zero to
 the pulse profile in an arbitrary way.
 The pulsar distance $d$ and the zero point difference $\Delta\phi_{\rm zp}$
 can be replaced in this equation with the fiducial phase.
 The definition of $\phf$, as articulated above, is 
\begin{equation}
 \phf = \phobs\left(\phmu\negthinspace=\negthinspace0, r\negthinspace=\negthinspace0\right) = \Omega d/c +
 \Delta\phi_{\rm zp} 
\end{equation}
 (from eq.~\ref{phobs1}),
 which can be inserted back into (\ref{phobs1}) to obtain:
 \begin{equation}
 \phobs = \phmu -
 \Omega \vec r \cdot \nobs / c + \phf \approx \phmu - r/\rlc + \phf,
 \label{phobs2}
 \end{equation}
 where on the right hand side I make the small-angle approximation: 
 $\vec r \cdot \nobs
 \approx r$ and use $\Omega/c = 1/\rlc$.
Remembering that $\phmu = \Omega t_e$ we get:
\begin{equation}
\Omega t_e = \phobs - \phf + r/\rlc
\label{omegate}
\end{equation}
which should be used in equations (16) and (17) of BCW in place of
`$\Omega t$' (the common practice of replacing $\Omega t$
in eqs.~(16) and (17) of BCW with $\phobs$ results in
PA curves that lag the fiducial phase 
by $3r/\rlc$ instead of the actual $2r/\rlc$; 
therefore they underestimate $r$ by a factor of $1.5$.)

Thus, the equation for the altitude-dependent PA curve reads:
\begin{displaymath}
\psir\approx\tan^{-1}\left[\frac{3(r/\rlc)\sin\zeta -
\sin\alpha\sin(\phobs - \phf + r/\rlc)}
{\sin\beta + \sin\alpha\cos\zeta(1-\cos[\phobs - \phf +
r/\rlc])}\right] +
\end{displaymath}
\begin{equation}
\hphantom{\psir\approx} + {\psira},
\label{psir1}
\end{equation}
where $\beta = \zeta - \alpha$ is the observer's `impact' angle.
The equation was obtained directly from eq.~(16) in BCW (in addition to
the use of $\Omega t=\phobs -\phf + r/\rlc$ I have changed the BCW's 
sign of the arctan term to agree with the observers' convention). 

The same approach may be applied to eq.~(17) of BCW (with $\phz = 3r/\rlc$)
to get:
\begin{displaymath}
\psir \approx \tan^{-1}\left[\frac{-\sin\alpha\sin(\phobs - \phf - 2r/\rlc)}
{\sin\beta + [1 - \cos(\phobs - \phf - 2r/\rlc)]\cos\zeta\sin\alpha}\right] +
\end{displaymath}
\begin{equation}
\hphantom{\psir \approx}+ \frac{10}{3}\frac{r}{\rlc}\cos{\alpha} + \psira,
\label{psir2}
\end{equation}
which is equivalent to (\ref{psir1}) within the accuracy of the BCW method,
ie.~up to the order of $r/\rlc$. The manually-added term 
$(10/3)(r/\rlc)\cos(\alpha)$
represents the vertical shift of the PA found by HA (the shift
given in HA refers to the clockwise definition of PA, and, therefore,
has opposite sign). For the polar-current flow that is close to the 
Goldreich-Julian value, this term is cancelled and can be neglected
in eqs.~(\ref{psir2}) and (\ref{psir}). In such a case, however,
the current-induced shift of PA must also be subtracted from eq.~(\ref{psir1})
for consistency (see Sect.~\ref{currents} below).

Yet another, most direct method to derive the equation for
the altitude-dependent PA curve is the following.
The center of the pulse profile precedes in phase the center
of the PA swing by $4r/\rlc$, ie.~$\phpa \approx \phprof + 4r/\rlc$.
The profile center itself precedes the fiducial phase $\phf$
by $2\rn$ (one $r/\rlc$ for the aberration and another $r/\rlc$
for the propagation time, see
eg.~DRH for details). Therefore,
the center of the PA curve lags the fiducial phase $\phf$ by
$4\rn-2\rn=2\rn$ and at any phase $\phobs$ the position angle $\psir$
which includes the rotational effects is equal to the `nonrelativistic' PA
(given by eq.~\ref{orig}) taken at the earlier phase $\phobs-2\rn$.
Thus, $\psir(\phobs) = \psi(\phobs\negthinspace-\negthinspace2\rn)$:
\begin{displaymath}
\psir \approx \tan^{-1}\left[\frac{\sin\alpha\sin(\phobs-\phf-2\rn)}
{\cos(\phobs-\phf-2\rn)\cos\zeta\sin\alpha
-\cos\alpha\sin\zeta}\right] +
\end{displaymath}
\begin{equation}
\hphantom{\psir \approx} + \frac{10}{3}\frac{r}{\rlc}\cos{\alpha} + \psira
\label{psir}
\end{equation}
This equation was simply obtained by replacing $\phobs$ in eq.~(\ref{orig})
with $\phobs-2r/\rlc$ and by replacing $\phz$ with $\phf$.
Simple trigonometry shows that it is equivalent to
eq.~(\ref{psir2}).

In the limit of $r \ll 10^{-2}\rlc$ all eqs.~(\ref{psir1}) -- (\ref{psir})
reduce to eq.~(\ref{orig}) and the centers of PA the curve, the pulse profile,
and the fiducial phase coincide.
In eqs.~(\ref{psir1}) -- (\ref{psir}) the radiation
electric field $\vec E_w$ is assumed to be along, rather than orthogonal to,
the direction
of electron acceleration $\vec a$ (in eq.~\ref{orig} $\vec E_w$
is assumed to be along
the magnetic field $\vec B$, or at least in the plane of a $\vec B$-field line).
For the case $\vec E_w \perp \vec a$ (or for
$\vec E_w \perp \vec B$), the value of PA given by eqs.~(\ref{orig}),
(\ref{psir1}), (\ref{psir2}), and (\ref{psir}) may need to be increased
by $90^\circ$ (as in the case of the Vela pulsar, Lai et al.~2001, Radhakrishnan
\& Deshpande 2001).
Obviously, if the projection of the rotation axis at the plane of the sky
is unknown
the terms $\psira$ in eq.~(\ref{orig})
and (\ref{psir1}) -- (\ref{psir})
must be considered free parameters to be determined from the fit to the data.
If $r$ is independent of pulse phase and we are not interested in the value of
$\psi_\Omega$, the term $(10/3)(r/\rlc)\cos{\alpha}$ can be merged with
$\psi_\Omega$ into a single fitting parameter. 

\begin{figure}
      \centering
      \includegraphics[width=0.5\textwidth]{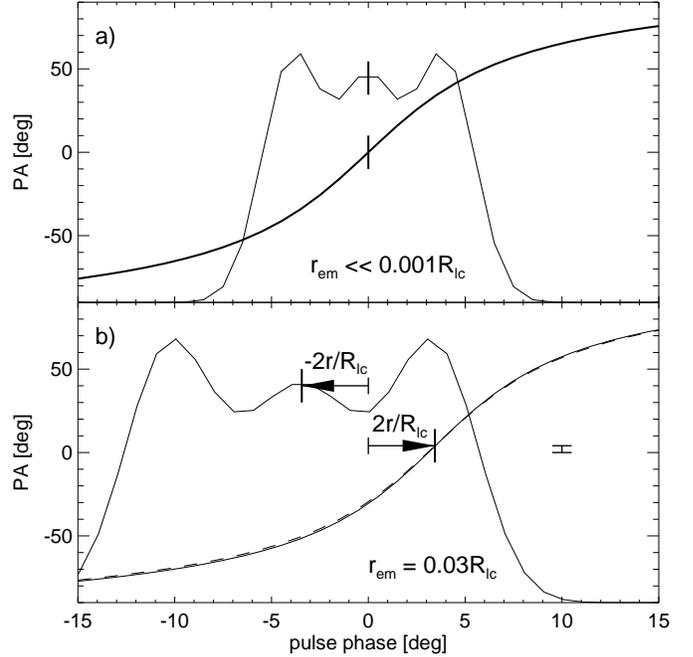}
      \caption{Influence of increased emission altitude on the observed
          PA curve. Panel {\bf a)} shows the RC69 case
          of negligible emission altitude. In {\bf b)} the emission altitude
	  has been rised up to $0.03\rlc$. As a result, both the intensity
	  profile and the PA curve move in opposite directions from 
	  the fiducial phase by the same angle of $2r/\rlc$ rad.
	  In both panels pulse phase zero is the fiducial phase
	  (`dipole axis phase'). 
	  In BCW, the only figure that has the fiducial phase at $\phobs=0$
	  is their fig.~3. The apparently small bar at $\phobs = 10^\circ$ 
	  shows
	  the vertical upward shift of the PA curve by $(10/3)(r/\rlc)
	  \cos{\alpha}$. The solid PA curve in {\bf b)} presents 
	  the approximate equation of BCW (eq.~\ref{psir1} in this paper).
	  The dashed line is just the appropriately shifted RVM curve 
	  (eq.~\ref{psir2}). In {\bf a)} they coincide. In the figure
	  $\alpha=45^\circ$ and $\zeta=41.3^\circ$. In all figures 
	  $\psi_\Omega=0$.
      }
      \label{fidpic}
   \end{figure}

Eq.~(\ref{psir}) depends only on the term $-\phf - 2r/\rlc$ which makes
it difficult to separate $\phf$ from $r$.
A special case when it is possible is when $r$ does not change across
the pulse profile.
In such a case the center of the PA curve
(the inflection point)
is located at the phase
\begin{equation}
\phpa \approx \phf + 2\rn.
\label{phpa}
\end{equation}
The center of the pulse profile is then located at
\begin{equation}
\phprof \approx \phf - 2\rn
\label{phprof}
\end{equation}
so that the shift of the PA with respect to the profile is $\phpa - \phprof
\approx 4\rn$.
Then (i.e.~for $r = const$) eqs.~(\ref{phpa}) and (\ref{phprof}) tell us that
\begin{equation}
\phf = (\phprof + \phpa)/2,
\label{phf1}
\end{equation}
ie.~$\phf$ is half way between the center of the pulse profile
and the center of the PA curve (Fig.~\ref{fidpic}).
If the nature and the quality of the data allow us to determine
$\phprof$ and $\phpa$, the value of $\phf$ can be calculated from (\ref{phf1})
under the asumption that $r = const$. In this specific 
case $\phpa$ can be determined
by fitting the \emph{non}relativistic PA curve (eq.~\ref{orig}) 
to the observed data.

\subsection{Relation between the radial distance of radio emission
and the pulse phase at which it is detected}

In the case of emission altitude that changes gradually 
with pulse phase $\phobs$, the eqs.~(\ref{psir1}) -- (\ref{psir})
must be supplemented by an analytical equation for $r=r(\phobs)$
to be useful.

If the distribution of emissivity in the CF frame
is symmetrical with respect to the dipole axis, the radial distance of
the radio emission $r$ is a function of only the angle 
$\rho$
between the emission direction in the CF and the dipole axis.
Let us assume that the emission region can be described by a simple
function $r(\rho)$ which can be inverted\footnote{If
more than one value of $r$ correspond to the same value of $\rho$
(ie.~if there are several layers of emission located above each other) 
one can separate
$r(\rho)$ into a few functions $\rho$, each of which is reversible.
Since emission from different layers can be observed simultaneously,
the Stokes parameters must be used as eg.~in KD83
or Mitra \& Seiradakis (2004).
}
into $\rho(r)$.

Fig.~\ref{emigeom} shows relative orientations of the dipole axis $\mu$,
and the CF-emission direction $\vec k^\prime$ at the moment 
when the radiation is directed towards the observer.
In the IOF it propagates along $\vec k \parallel \nobs$.

The radiation from some point at radial distance
$r$ is directed towards the observer only when $\vec \mu$ is rotated by
an appropriate angle $\phmu=\phemab=\Omega t_{\rm em}$. Because the aberration
advances the radiation by $\rn$ in phase,
the angle $\phemab$ is smaller by $\rn$ than the angle 
$\phem$
by which the dipole would need to be rotated
in the absence of the aberration 
to become aligned with the line of sight (see Fig.~\ref{emigeom}):
\begin{equation}
\phem\simeq\phemab + \rn = \Omega t_{\rm em} + \rn.
\label{phem}
\end{equation}

From (\ref{omegate}) and (\ref{phem}) we learn that the radiation 
is shifted toward earlier phases by $2\rn$ with respect to $\phf$
and is detected at the phase:
\begin{equation}
\phobs \simeq \phf + \phem - 2\rn.
\label{phobs}
\end{equation}
By substituting $\phobs$ in eq.~(\ref{psir}) by the above formula,
and using $\phem\simeq\phemab+\rn$, as well as $\phemab=\Omega t_{\rm em}$
one can easily verify that eq.~(\ref{psir}) is equivalent with eq.~(17) in
BCW.

Unlike $\phemab$ (see Fig.~\ref{emigeom}), the azimuth $\phem$ 
of the non-aberrated emission direction
$\vec k^\prime$ is associated with the angle $\rho$ between $\vec \mu$ and
$\vec k^\prime$ through:
\begin{equation}
\cos\left[\rho(r)\right] =
\cos\phem\sin\alpha\sin\zeta^\prime + \cos\alpha\cos\zeta^\prime,
\label{rho}
\end{equation}
which is a direct form of the spherical trigonometry cosine theorem
applied for the triangle $(\vec \Omega, \vec k^\prime, \vec \mu)$.
Calculating $\phem$ from (\ref{rho}) and inserting into
eq.~(\ref{phobs}) gives
\begin{equation}
\phobs(r) \simeq \pm\cos^{-1}\left(\frac{\cos(\rho(r)) - \cos\alpha\cos\zeta}
{\sin\alpha\sin\zeta}\right) + \phf - 2\frac{r}{\rlc},
\label{phobs3}
\end{equation}
where we ignored the insignificant difference between $\zeta$ and 
$\zeta^\prime$. The `$+$' sign at the arccos term corresponds to the trailing
whereas the `$-$' sign to the leading part of the open field line region.

Since the argument of the function $\arccos$ in eq.~\mref{phobs3}
does not have to be
small (e.g.~for small dipole inclination $\alpha$ the profile width
can reach several tens of degrees), 
the equation in general cannot be inverted to obtain
a simple analytical formula for $r(\phobs)$. To use eq.~(\ref{psir})
for phase-dependent $r$, the latter needs to be determined from 
(\ref{phobs3}) numerically.

\subsection{Numerical example}

Eq.~\mref{phobs3} is useful for a quick examination of the shape of PA curves 
for a variety
of emission regions, ie.~for various functions $r(\rho)$.
A convenient way of doing this is to define a dense table of angles
$\rho_i$, and calculate the corresponding vector of $r_i(\rho_i)$ for some 
chosen
function $r(\rho)$ (or vice versa: to define
$r_i$ and calculate $\rho_i(r_i)$
for arbitrarily selected $\rho(r)$). 
These are next used in eq.~\mref{phobs3} to calculate
the table of the corresponding values of $\phobsi$. 
The tables $r_i$,
and $\phobsi$ can then be directly used in eq.~\mref{psir1} 
(or \ref{psir1bis}) to calculate the shape of the PA curve $\psi(\phobs, r)$.

   \begin{figure}
      \centering
      \includegraphics[width=0.5\textwidth]{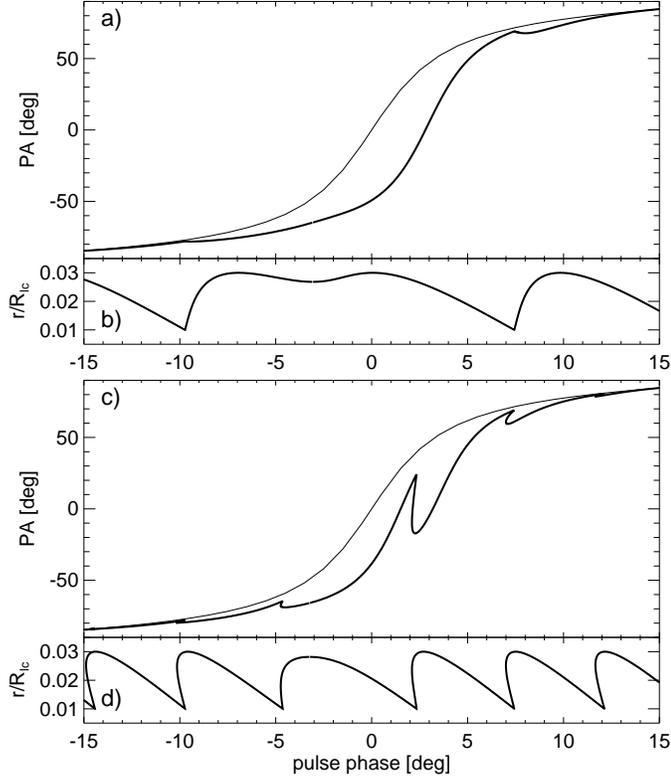}
      \caption{Altitude-dependent PA curves (thick lines in {\bf a} and {\bf c})
         calculated by using eqs.~\mref{phobs3} and \mref{psir1}
	 for the emission region of eq.~\mref{alts} 
	 with $a_0=2^\circ$ (top) 
	 and $a_0=1^\circ$ (bottom).
	 The thin line shows the standard
	 RVM curve for reference. Panels {\bf b} and {\bf d} 
	 show the corresponding
	 radial distances in units of $\rlc$. Phase $\phobs=0$ 
	 is the fiducial phase and I used $\alpha=45^\circ$ and 
	 $\zeta=43^\circ$ in the figure.
      }
      \label{altsPA}
   \end{figure}

Fig.~\ref{altsPA} presents an example of such procedure performed for
the emission region given by:
\begin{equation}
\frac{r}{\rlc} = 0.01 + 0.02\left|\sin\left(\frac{\rho}{a_0}\right)\right|,
\label{alts}
\end{equation}
where the constant $a_0=2^\circ$ was used in panels a and b, whereas 
$a_0=1^\circ$ in c and d (this somewhat strange shape was chosen to
make the distortions of the PA curves easily noticeable by eye).
The curves $r(\phobs)$ (as given by eq.~\ref{phobs3}) are shown in panels
b and d. One can see the characteristic skewing of the $r(\phobs)$
function toward early phases, which is caused by the APT effects.
In panel d, the radial distance $r$ increases with phase so fast 
that the radiation from the upper parts
of the emission region overlaps in phase with radiation emitted
at lower altitudes. The corresponding PA curve (panel c) is not 
a mathematical function of phase (there are a few regions with
three values of PA referring to the same $\phobs$). To look normal 
it needs to be cut into pieces and Stokes-summed
(for details see Mitra \& Seiradakis 2004; KD83).

\section{Limitations on the validity and applicability of the linear theory}
\label{limits}

   \begin{figure*}
      \includegraphics[width=0.8\textwidth]{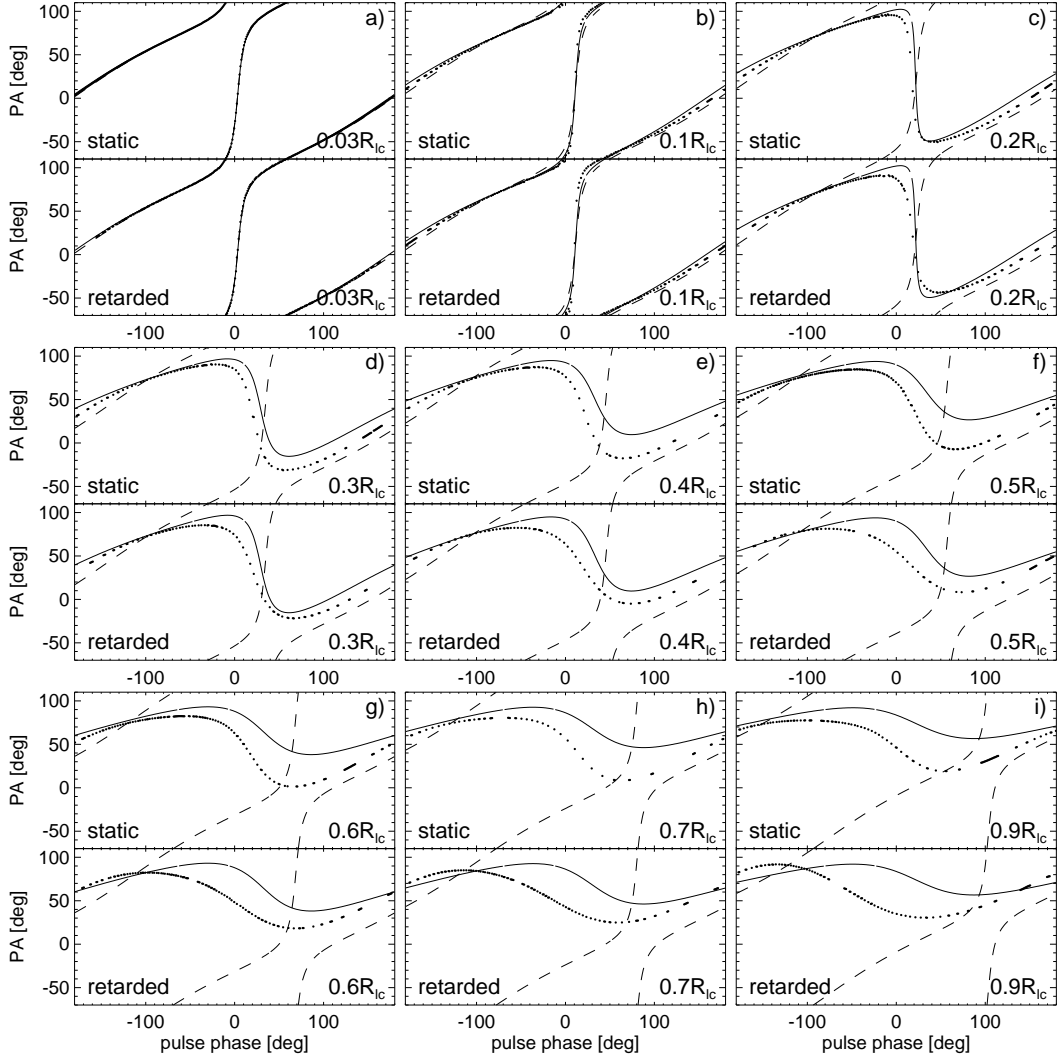}
       \caption{Variations of polarization angle curve due to 
                increasing emission altitude, marked in the bottom right
		corners of each panel. The dots, often merging into a 
		thick solid line, present exact numerical solution.
		The thin solid line presents the equation (\ref{psir1}).
		The dashed line is for eqs.~(\ref{psir2}) and/or (\ref{psir}).
		For each altitude the numerical curve is shown for 
		both the static-shape dipole (letter-marked panels) and the 
		rotationally-distorted vacuum dipole (below the static case).
		The figure was calculated for $\alpha=45^\circ$, 
		$\zeta=41.3^\circ$, $\phf=0$, and $\psira=0$.
       }
      \label{pacurves}
      
   \end{figure*}

The equations of Sect.~\ref{pacurv} refer to phase shifts that are usually tiny 
and can be easily affected by several interfering effects discussed 
in this section. 
Moreover, the accuracy of the analytical results of previous 
sections is limited by their inherent error of $(r/\rlc)^{1/2}$. 
The validity of the analytical theory is therefore severly limited:
on one hand by the too small magnitudes of the phase shifts
(too small to be reliably measured, or to be unaffected by
the disturbing effects); on the other hand by the poor accuracy of the theory
at larger $r$.
The following subsections are to show that it makes practically
no sense to apply the theory for $r\ga 0.1\rlc$ ($\Delta\phobs \ga
10^\circ - 20^\circ$).

\subsection{Asymmetry of the open field line region caused by rotational 
distortions of the magnetic dipole (sweepback)}

The strict symmetry of the open volume with respect to the $\om$-plane 
is unlikely, and it has been shown (Dyks \& Harding 2004, hereafter DH04) that
even very small effects (that at low altitudes are of high order in $r/\rlc$) 
can easily introduce
large asymmetry.
Here I discuss the rotational deformations of the vacuum dipole
(DH04; Shitov 1983, Deutsch 1955),
because they are easy to analyze (thanks to the existing
analytical formulae) and they seem to be qualitatively similar
to those obtained in recent plasma-loaded simulations (eg.~Spitkovsky 2008).
As we show in detail in DH04, in the lowest order 
(of $(r/\rlc)^2$, which is already higher than the $r/\rlc$ 
accuracy of equations in the preceding sections)
the rotational sweepback does not introduce any asymmetry to the $B$-field 
direction. 
The asymmetric change of $\vec B$-field direction is of the order 
of $(r/\rlc)^3$
at low altitudes (completely negligible \emph{there}).
Despite this, a very strong asymmetry of the open volume
(of magnitude $\sim(r/\rlc)^{1/2}$) is generated by the sweepback.
This is because the low-altitude outer boundary of the open volume 
is determined
by the geometry of $\vec B$ \emph{at the light cylinder}, where the sweepback
becomes a strong effect (first order, in the sense that there 
we have $(r/\rlc)^a \sim 1$, regardless of $a$).
This strong distortion is traced back towards low altitudes
through the continuity of magnetic field lines, and results in strong
(much stronger than $(r/\rlc)^3$) asymmetry of the open volume.

This effect can be taken into account by replacing the delay-radius relation
(eq.~\ref{bcw}) with the `misalignment' formula of DH04:
\begin{equation}
\Delta\phi \simeq 4\frac{r}{\rlc} - F\left(\frac{r}{\rlc}\right)^{1/2},
\label{misal}
\end{equation}
where $F$ depends on $\alpha$ and $\zeta$ and has typically
the magnitude of $\sim0.2$. 
Eq.~\mref{misal} differs from the original delay-radius relation
only in that it takes into account the azimuthal asymmetry of the 
outer boundary of the open field line region.
However, I do not refer to eq.~(\ref{misal}) with 
the name `delay-radius' relation because the shift of position angle curve
does not have to be the positive delay anymore -- it can now \emph{slightly} 
precede the profile midpoint, see DH04 for more details.

Those, who believe that the \emph{rotating} vacuum dipole is a better
approximation of the real $\vec B$ than the static vacuum dipole 
should use
eq.~(\ref{misal}) instead of the original delay-radius relation of BCW.
What may be most reasonable is to use both 
the eq.~\mref{misal} and \mref{bcw} as
the lower and upper limit for $r$, respectively.

\subsection{Current-induced distortions of magnetic field}
\label{currents}

The magnitude of the current-induced distortions of magnetic field
can be easily estimated with a simplification of Ampere's law
$\nabla\times \vec B = c^{-1}(4\pi \vec J +\partial \vec E/\partial t)$ 
into the crude form of $\Delta B/L \sim J/c$, where $\Delta B$ is the
magnetic field generated by the current density $J$, and $L$ is 
the characteristic scale. For longitudinal polar currents
of the Goldreich-Julian (GJ) magnitude we have 
$J \sim J_{\rm GJ} \sim \rho_{\rm GJ}c$, where
$\rho_{\rm GJ}\simeq\Omega B/(2\pi c) \sim B/\rlc$ is the 
GJ density, whereas the transverse scale 
$L \simeq r\thlo \simeq r(r/\rlc)^{1/2}$
corresponds to the open field line region.
The distortions of $B$ are therefore of the order of 
$(\Delta B/B)_{\rm polar} \sim 
(r/\rlc)^{3/2}$. For toroidal currents (due to the corotation of 
$\rho_{\rm GJ}$) we have $J \sim \rho_{\rm GJ}v_{\rm rot}$ where 
$v_{\rm rot}=\Omega r=cr/\rlc$ is the local corotation velocity, and $L \sim r$.
This results in smaller distortions of 
$(\Delta B/B)_{\rm trdl}\sim (r/\rlc)^2$.

HA find that the polar currents shift the PA curve 
downward by $\Delta\psi_J = -(10/3)(r/\rlc)(J/J_{\rm GJ})\cos{\alpha}$.
A spectacular way of estimating the current density
has recently become possible thanks to the observations of part-time pulsars
(Kramer et al.~2006) and suggests that $J\sim J_{\rm GJ}$
during the `on' phase of pulsar emission.

The simplest way to include the effects of current in the approximate
eqs.~\mref{psir1} -- \mref{psir} is to subtract the term
$(10/3)(r/\rlc)(J/J_{\rm GJ})\cos{\alpha}$ with $J/J_{\rm GJ} = 1$
from their right-hand sides,
to obtain:
\begin{displaymath}
\psir \approx \tan^{-1}\left[\frac{3(r/\rlc)\sin\zeta -
\sin\alpha\sin(\phobs - \phf + r/\rlc)}{\sin\beta + \sin\alpha\cos\zeta(1-\cos[\phobs - \phf +
r/\rlc])}\right] +
\end{displaymath}
\begin{equation}
\hphantom{\psir \approx}
- \frac{10}{3}\frac{r}{\rlc}\cos{\alpha}\left(\times\frac{J}{J_{\rm GJ}}
\right) + {\psira}
\label{psir1bis}
\end{equation}
for the BCW formula of eq.~\mref{psir1}, and
\begin{displaymath}
\psir \approx \tan^{-1}\left[\frac{\sin\alpha\sin(\phobs-\phf-2\rn)}
{\cos(\phobs-\phf-2\rn)\cos\zeta\sin\alpha
-\cos\alpha\sin\zeta}\right] +
\end{displaymath}
\begin{equation}
\hphantom{\psir \approx} + \psira
\label{psirbis}
\end{equation}
for the `shifted RVM' curve of eqs.~\mref{psir2} and \mref{psir}.

\subsection{Shapiro delay}

To make things possibly simple,
the importance of general relativistic time delays has been neglected so far.
These should not affect the delay-radius relation, because
it refers only to a single $r$, and is insensitive to 
any time delays (whether gravitational or flat spacetime).
They do affect, however, all formulae which refer to altitude
differences (as do all the equations that explicitly 
involve both $r$ and $\phf$).
Let us then use the Schwarzschild metric to learn when the Shapiro delay 
cannot be neglected.
The time for a ray to travel radially in the gravitational field
from some $\rmin$ up to $\rmax$ is
\begin{equation}
\Delta t_{\rm Schwrzld} = \int_{\rmin}^{\rmax}\frac{dr}{c(1 - r_g/r)},
\end{equation}
where $r_g=2GM_{NS}/c^2$. Instead of that, our flat spacetime formulae 
assume the time is the R\"omer delay $(\rmax-\rmin)/c$, 
so they miss the fact that 
the low altitude emission components (emitted at $\rmin$) are additionally
Shapiro-delayed by
$\Delta t_{\rm Sh} = \Delta t_{\rm Schwrzld} - (\rmax-\rmin)/c$ with respect
to the high-altitude emission components emitted at $\rmax$.

In the case of millisecond pulsars (MPs) 
this effect can become easily measurable:
for $\rmin = 10^6$ cm, $\rmax = 0.2\rlc$, and $P=3$ ms, we obtain
$\Delta t_{\rm Sh} = 2\cdot 10^{-5}$ s which corresponds to the phase shift
of $2.48^\circ$. For the same $\rmin$ and $\rmax/\rlc$ but for $P=1.5$ ms
the phase shift is $1.95^\circ$. It become slightly smaller because 
for the fixed $\rmax/\rlc=0.2$ the altitude difference decreased
(for $P=3$ ms $\rlc$ is $14.3 \cdot 10^6$ cm, whereas for $P=1.5$ s 
it is twice smaller).
Note that in the estimate I take $\rmax=0.2\rlc$ as a blind guess of 
the upper limit for radio emission regions in millisecond pulsars. 
This is simply because the known estimates or radio emission 
altitudes for classical pulsars (a few tens of $\rns$, DRH;
Gupta \& Gangadhara 2003; Kijak \& Gil 1997) would locate
the emission region beyond the light cylinder in the millisecond pulsars.
If one takes $\rmax = \rlc$, the gravitational phase shifts become
$5.44^\circ$ and $8.41^\circ$ for $P=3$ and $1.5$ ms, respectively.
Thus, the equations that rely on altitude differences 
cannot be reliably applied to fast millisecond pulsars
(in particular, it is not safe 
to use eq.~(\ref{phprof}) to determine emission altitudes from
the core-cone shift).
Another reason for not applying the approximate theory to the MPs 
is that it is very inaccurate for altitude differences of $\Delta r 
\ga 0.1 \rlc$, which are likely in fast MPs 
even if the radio emission
extends radially by only one $\rns$. To calculate the 
phase shifts for MPs even more accurately, one would have to use the metric 
for a fast rotating neutron star (eg.~Braje et al.~2000; Gonthier \& Harding
1994; see also
D'Angelo \& Rafikov (2007) for a detailed description of how the altitude
differences affect timing).

In the case of classical pulsars (with $P\sim 1$ s and $\rmax \simeq 50\rns$),
the Shapiro delays can be safely ignored, because they comprise a tiny 
fraction of the (now longer) rotation period. The phase shifts typically
have a few hundredths of a degree.

\subsection{Limited accuracy of the lowest-order theory}
\label{accuracy}

The relativistic phase shifts are of the order of $r/\rlc$, whereas
the lowest-order terms that have been neglected in derivation of
eqs.~\mref{psir1} -- \mref{psir} have magnitude of $\sim (r/\rlc)^{3/2}$.
This means that the analytical approximations for the polarization angle
have a fractional error of $100(r/\rlc)^{1/2}$ percent, which is as large as
$30$\% already at $0.1\rlc$.
The approximate equations \mref{psir1} -- \mref{psir}
are compared to the exact numerical PA curves in Fig.~\ref{pacurves}. 
The numerical PA curves were calculated for fixed emission altitudes
(shown in bottom right corners) and only involve the kinematic effects
of corotation (the method of calculation is described in sect.~2 of 
Dyks et al.~2004a).
It is seen that for any $r \ga 0.1\rlc$ there are considerable
differences between the exact result (dots) and the approximate formulae.
One can also see that at large altitudes 
the approximate formula of BCW (eq.~\ref{psir1}, thin solid lines) 
performs much 
better than the appropriately shifted classic equation of Komesaroff
(eqs.~\ref{psir2} and \ref{psir}, dashed lines).

An interesting effect seen in Fig.~\ref{pacurves} is that for large altitudes
the numerical
PA curves strongly tend to assume the distorted-sine-like 
(`equatorward') shape, in spite of that the figure is for the poleward
viewing geometry ($\alpha=45^\circ$, $\zeta=41.3^\circ$).
The range of $\zeta$ with the equatorward PA shape is not limited to the
range $(\alpha, \pi-\alpha)$ and is instead increasing with altitude.
For the specific case shown in Fig.~\ref{pacurves} 
the numerically determined range of the equatorward PA curve
is $(43^\circ,\pi-43^\circ)$ for $r=0.1\rlc$, and $(38.5^\circ,\pi-38.5^\circ)$
for $r=0.2\rlc$ (with numerically-limited accuracy of $\sim0.5^\circ$).
The BCW equation for the PA (eq.~\ref{psir1}) does reproduce this
behaviour (to some limited degree), but the fixed-shape PA curve
of the `relativistically shifted' RC69 model
(eq.~\ref{psir2} or \ref{psir})
fails to do this, which results in the large disagreement
already visible in panel c of Fig.~\ref{pacurves} (for $r=0.2\rlc$).

Another interesting effect in Fig.~\ref{pacurves} can be seen 
if one compares PA curves of the static-shape dipole to the distorted
(`retarded') dipole case. At large altitudes ($r \ga 0.7\rlc$)
the sine-like PA curves for the distorted dipole (Fig.~\ref{pacurves}h,i)
move leftward (toward earlier phases)
and have their steepest gradient point shifted leftward with respect
to the fiducial phase (opposite than expected from the BCW theory).
This effect is caused by backward bending of magnetic field lines
(the sweepback) which shifts the spatial location of ``magnetic axis"
(here I mean a locally straight $\vec B$-field line at a given $r$) 
foreward, into the leading part of pulsar magnetosphere.
At large $r$ this effect seems to dominate the straightening of electron
trajectories.

The distorted-sine shape of the high-altitude PA curves, with the steepest 
gradient point well ahead of the `dipole-axis' phase, bears
close resemblance to the unusual PA curve of the pedestal radio emission
of B1929$+$10, where the steepest gradient \emph{precedes} the
main pulse by as much as $18^\circ$ (Rankin \& Rathnasree 1997;
Everett \& Weisberg 2001). 
This interpretation would imply
that the pedestal radio emission originates from $r\sim0.7\rlc$.
However, below I also mention other possible interpretations of this effect.

\subsection{Limitations due to the specific radio emission mechanism
assumed in the theory}
\label{noncurv}

The radio emission mechanism assumed in the BCW theory is the curvature 
radiation caused by the `macroscopic' acceleration due to the 
curvature of magnetic field lines. The acceleration is small enough that
the corotation can modify it to produce the delay-radius relation of 
eq.~\mref{bcw}. The operation of the curvature radiation in pulsar 
magnetosphere is, however, still a matter of debate
(eg.~Luo \& Melrose 1992; Kunzl et al.~1998).

The other emission processes (eg.~the direct or inverse-Compton-scattered
plasma emission, or the synchrotron emission) typically involve
much larger accelerations due to microscopic motions,
that are unlikely to be noticeably affected by the corotation
(eg.~the acceleration due to gyration exceeds the macroscopic acceleration
due to the corotation by a factor $\omega_B/\Omega \sim 10^8[B/(10^6{\rm G})]
(\gamma/10^3)^{-1}$, where $\omega_B$ is the gyration frequency,
Takata et al.~2007).

It is then tempting to speculate that the PA curves for such emission 
processes would not be delayed by the BCW effect. But the aberration and
propagation time effects would still be there, and would shift the PA
curve by $2r/\rlc$ toward \emph{earlier} phase than the main pulse.
This may be a mechanism responsible for the leftward shift of the pedestal
PA in B1929$+$10. In Dyks, Rudak \& Rankin (2007; hereafter DRR) we provide
other arguments for non-curvature origin of the pedestal emission.
The main pulse emission, on the other hand, does not exhibit 
the abnormal shift
which makes it more consistent with the curvature radiation. 

\section{Anti-BCW shifts of PA curve}
\label{antibcw}

The best examples of this phenomenon are provided by the pedestal
radio emission of B1929$+$10
and the weak bridge of radio emission that connects the main pulse and 
interpulse of B0950$+$08. The PA curves of these extended components
(fitted \emph{without} the part under the main pulse, 
see Everett \& Weisberg 2001)
have their steepest gradient
points well on the leading side of the main pulse.
So far the following mechanisms have been considered for this effect:
\begin{enumerate}
\item The high-altitude curvature emission from roughly fixed altitude
(Sect.~\ref{accuracy} in this paper and Fig.~\ref{pacurves}h,i).
This probably cannot explain the case of B0950$+$08, for which the PA curve
has the poleward shape.
\item Radio emission mechanisms different than curvature radiation
(sect.~\ref{noncurv}). Possibly consistent with the parallel-ICS
interpretation of double notches (DRR).
\item Inward curvature radiation. Proposed simply as an inverted version
of the BCW effect (Dyks et al.~2005). The `pulsar shadow' model
of double notches, that has led to this interpretation, has been superceeded
by the model described in DRR.

\item Yet another interpretation can be devised using the `limiting 
polarization radius' idea: polarization of the radiation propagating
through the inner (dense) regions of the pulsar magnetosphere 
can follow the \emph{local} direction of magnetic 
field (eg.~Cheng \& Ruderman 1979; Melrose 1979; Barnard 1986) 
until the radiation reaches the polarization-limiting radius $r_{\rm pol}$.
The plasma-density-dependent estimates of $r_{\rm pol}$ are highly uncertain
(eg.~Lyubarsky 2002) but it should be significantly larger than the radial
distance of the emission region.

High-up in the pulsar magnetosphere the polar beam of radio emission,
emitted at low altitudes,
propagates through the \emph{trailing} part of open field line region 
(see fig.~1 in Dyks \& Rudak 2002, or fig.~1 in Lyubarsky 2002).
Therefore, the polarization imprinted in the beam at $r_{\rm pol}$
reflects the trailing part of the polarization curve, with the steepest
gradient point shifted considerably towards early phases.
The same mechanism was proposed by Lyubarsky (2002) to explain
the shallowness of PA curves observed in millisecond pulsars
(Xilouris et al.~1998).
Let me emphasize that the parallel-ICS model of pedestal emission
mentioned in point (ii) above, actually \emph{necessitates} 
this propagation-induced linear polarization. Otherwise the observed
polarization degree would be very low due to the convolution of many
micro-beams contributing to the line of sight.
\end{enumerate}

\section{Discussion -- aberration or not?}
\label{abornot}

It is shown in Sect.~\ref{simpderiv} that the magnitude of the shift given 
by the delay-radius relation has nothing to do with the aberration
(understood as the normal beaming effect).
On the other hand, Hibschman \& Arons mention a phase shift
of $3r/\rlc$ (their eq.~4) and claim that it is just the aberration
(``simple beaming") that is responsible for the remaining $r/\rlc$ 
part of the total delay.
Do we have a contradiction here?

Not really, though the wording of HA can easily 
be misunderstood. 
Both in the BCW and HA, the delay-radius relation is derived
as a difference of two moments of emission: 1) 
the moment when the radiation 
bearing
the steepest-gradient PA 
is directed to the observer (this happens at $\Omega t_e=3r/\rlc$)
and 2) an earlier moment when the
radiation in the middle of the profile (emitted in the 
$\om$-plane
in the CF) becomes directed toward the observer. 
HA explicitly notice that the latter does \emph{not} take place
at the moment $t_e=0$,
when the dipole axis is in the $\vec \Omega$-observer plane:
instead, because of the aberration it happens at $\Omega t_e=-r/\rlc$
(see Fig.~\ref{emigeom} and eq.~\ref{phem} with $\phem=0$).
The key point that is not mentioned in their description, however, 
is that the moment $\Omega t_e=3r/\rlc$ at which the steepest gradient 
radiation 
is directed toward the observer has already been also advanced in time
by the same aberration angle of $r/\rlc$. 
Thus, \emph{the aberration, understood as the normal beaming effect,
works identically both at the center of the main pulse, as well
as on its trailing side and contributes practically nothing to the shift
of the PA curve with respect to the intensity profile.}

Another issue related to aberration is mostly nomenclatural.
In this paper I describe the origin of the delay-radius relation as
the `straightening of electron trajectories' when 
they are transformed from CF to IOF. Or, one can say the effect is due to
the transformation of electron acceleration from the non-inertial CF to IOF.
In the wording of HA the phase shift caused by these effects is described
as `aberrational' too 
(see eg.~their Appendix F and G). An argument that can (possibly)
justify this is eq.~\mref{vel} (their eq.~F1). The addition of velocities
(aberration) given in this equation determines the electron velocity
in IOF, which is next differentiated to obtain IOF acceleration.
Of course, what really matters is not the fact that we add the corotation 
velocity, but the fact that the added velocity is time-dependent
(or that CF is non-inertial). Since the phenomenon of aberration by itself 
refers only to the velocity of a reference-frame, and not to its acceleration,
the term `aberration' misses the essence of the effect, 
which is the \emph{time-dependent 
nature} of the corotation velocity in \mref{vel}.

A word of comment on papers that have assumed that the PA curve
is shifted \emph{forward} in phase, just as the intensity profile does
(eg.~KD83; Xu et al.~1997; note that the work of KD83 
was published in the pre-BCW era).
If one corrects the analysis of KD83 for the direction of the PA shift,
the altitude order of their four emission regions (see.~their fig.~16)
should be turned upside-down, ie.~the region 4 should be at the top, 
region 3 below 4, etc.

Finally a note related to the high-energy emission.
Electrons with Lorentz factors $\gamma \simeq 10^6$ -- $10^7$
that move along the dipole axis trajectory with $\rhomu = \rlc/2$ and 
$P \sim 0.1$ s
emit curvature radiation that extends up to $0.1$ -- $100$ MeV, respectively.
This energy is too low for pair production (see fig.~7 in Dyks \& Rudak 2002).
However, this radiation may contribute to the observed X-rays and gamma-ray
pulse profiles and spectra. In the numerical models of the 
high-energy emission from pulsars
it is therefore necessary to calculate the curvature radiation
using the radius of curvature  of electron trajectory
in the inertial observer frame.

\section*{Acknowledgements}

I thank B.~Rudak for comments on the manuscript.
This paper was supported by the grant N203 017 31/2872
of the Ministry of Science and Higher Education.

\appendix

\section{Simple derivation of eq.~(1)}
\label{ladyb}

\begin{figure}
      \centering
      \includegraphics[width=0.5\textwidth]{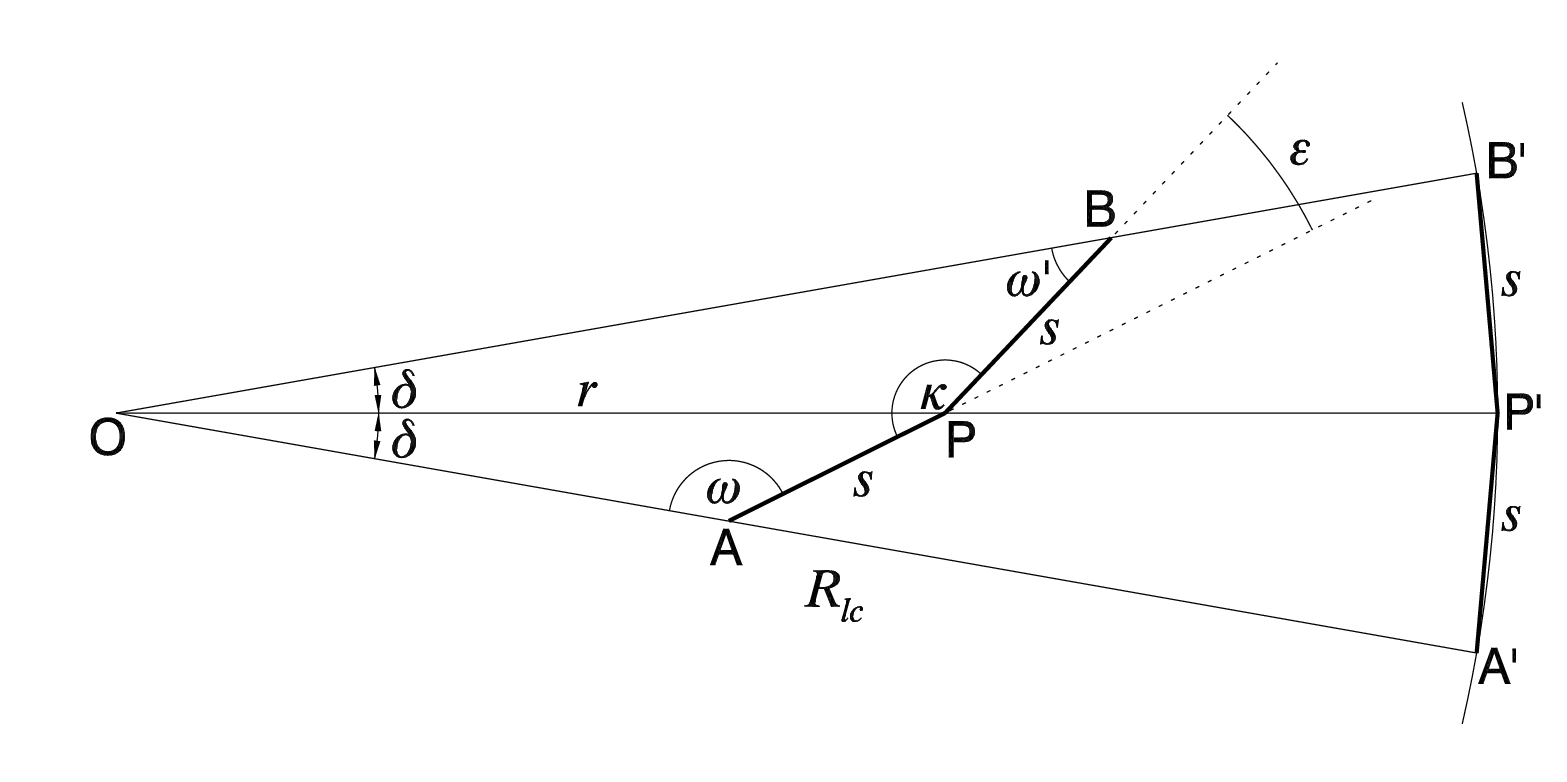}
      \caption{Trajectory APB of a relativistic electron that is
       moving along the straight dipole axis in the corotating frame. 
       Rotation axis is perpendicular
       to the page at O. The points A', P', and B' present locations
       of a point corotating at the light cylinder. The diagram
       is used in the school-style derivation of eq.~(\ref{rheq}) 
       (Appendix \ref{ladyb}).
      }
      \label{ladybird}
   \end{figure}

Let us consider the geometry shown in Fig.~\ref{ladybird}.
The three positions 
A', P', and B'
of a point at the light cylinder
are separated by the same path increment $s\simeq \rlc\delta$
so that the triangles OA'P' and OP'B' are identical. 
As soon as we mark in this figure some arbitrary position, 
say A, of an electron that is moving radially in the corotating frame,
the trajectory APB of the electron becomes uniquely determined.
This is because: 1) the electron also moves with the speed of light $c$
(as the point at the 
light cylinder does), so that we must have $|AP| = |PB| = s$ and 2)
the electron has the same angular velocity $\Omega$ so that
both AP and
PB must subtend their respective angles $\delta$.

The electron trajectory has a radius of curvature $\rhomu = s/\cal \scriptstyle
E$ (for elementary reasons), where ${\cal \scriptstyle E} = \pi - \kappa$ 
and $\kappa$ is the angle between AP and PB.
For the quadrangle OAPB we can write $2\delta + \omega + \kappa + 
\omega^\prime = 2\pi$ which implies ${\cal \scriptstyle E} = 2\delta +\omega
+\omega^\prime - \pi$. The sine theorem for the OPB triangle gives
$\sin\omega^\prime = r\sin(\delta)/s \simeq r\delta/s \simeq r/\rlc$,
i.e.~$\omega^\prime \simeq \arcsin(r/\rlc)\simeq r/\rlc$.
For the OAP triangle one has $\sin\omega = r\sin(\delta)/s\simeq
r/\rlc$ (same as for $\omega^\prime$) but this time $\omega$
is in the range $\pi/2 < \omega < \pi$
(see Fig.~\ref{ladybird}) so that $\omega = \pi - \arcsin(r/\rlc) \simeq 
\pi - r/\rlc$. Thus we have ${\cal \scriptstyle E} \simeq 2\delta\simeq 
2s/\rlc$
so that $\rhomu = s/{\cal \scriptstyle E} \simeq\rlc/2$. 

A formal derivation of eq.~(\ref{rheq}) assumes that the electron velocity
is given by:
\begin{equation}
\vec v \simeq c\hat b + \vec \Omega\times \vec r,
\label{vel}
\end{equation}
where $\hat b$ is a unit vector along the local direction of $\vec B$,
and the tiny difference between the electron speed along $\hat b$
and $c$ is neglected (hence the `$\simeq$' sign).
The electron acceleration is then:
\begin{eqnarray}
\vec a & = & \frac{d\vec v}{dt} = c\frac{d\hat b}{dt} +
\vec \Omega\times \frac{d\vec r}{dt} = c\frac{d\hat b}{dt} + \vec \Omega\times
\vec v = \label{a}\\
& = & c\frac{\partial \hat b}{\partial t} +
c\left(\vec v \cdot \nabla\right)\hat b + c\vec \Omega\times\hat b
+ \vec \Omega\times(\vec\Omega\times\vec r),\label{aa}
\end{eqnarray}
where the Lagrange derivative of $\hat b$ in (\ref{a})
has been expressed explicitly in (\ref{aa})
and $\vec v$ from (\ref{vel}) has been inserted into the last term of (\ref{a}).

The factor $\partial\hat b/\partial t$ represents the rotation 
of essentially radial unit vector
$\hat b$ with angular velocity $\Omega$ and therefore 
the first term has the magnitude of $\Omega c$.
The second term in eq.~(\ref{aa}) represents the contribution 
to acceleration that
results from curvature of
magnetic field lines 
and is zero at the dipole axis
(solely this term fully represents
the original model of the RC69).
The third term has the same magnitude as the first one: $|c\vec 
\Omega\times\hat b| \simeq \Omega c$. 
The last term has the magnitude of $\Omega$ times the local
corotation velocity $\vec v_{\rm crt} =\vec \Omega\times \vec r$
and it is negligible at low altitudes where $v_{\rm crt} \ll c$.
Thus, $|\vec a|\simeq 2\Omega c$ and $\rhomu \simeq c^2/a \simeq 
\rlc/2$.

For $\alpha\ne 0$ (and $\zeta=\alpha$)
the two non-zero terms in eq.~\ref{aa} (and therefore $\vec a$)
are smaller by a factor of $\sin{\alpha}$, so that $\rhomu$
is larger by $1/\sin{\alpha}$. The zero-curvature regions
are located closer to the dipole axis by the $\sin{\alpha}$ factor.
However, the delay-radius relation retains its usual form
($\alpha$-independent), because the
$\sin{\alpha}$ factor is cancelled by the `not a great circle' effect.

Yet another derivation of eq.~(\ref{rheq}) can be found in Thomas \& Gangadgara
(2005) (see their eq.~29).

\section{Aberration and propagation time delays}
\label{abret}


The aberration is understood as the change of emission direction
of radio waves caused by the transition from the Lorentz frame 
instantaneously comoving 
with a corotating emission point to the inertial observer frame (IOF).
I assume that in the corotating frame (CF) the radio waves are emitted 
in the direction $\vec k^\prime$ which is tangential to the
direction of dipolar magnetic field ($\vec k^\prime = \pm \vec B/|\vec B|$).
Then, in the IOF the emission occurs
in the direction:
\begin{equation}
\vec k = \frac{\vec k^\prime + [\gamma_{\rm crt} + (\gamma_{\rm crt}-1)
(\vec \beta_{\rm crt} \cdot \vec k^\prime ) / \beta_{\rm crt}^2
]\vec \beta_{\rm crt}}{\gamma_{\rm crt}(1+ \vec \beta_{\rm crt} \cdot \vec k^\prime)},
\label{aberration}
\end{equation}
where $\gamma_{\rm crt}=(1-\beta_{\rm crt}^2)^{-1/2}$, and $\vec \beta_{\rm crt}
\equiv \vec v_{\rm crt}/c=\vec \Omega\times\vec r/c$.
Eq.~\mref{aberration} is just the general Lorentz
transformation for an arbitrary velocity vector $\vec v$, 
in which I substituted
$\vec k = \vec v/c$ and $\vec k^\prime = \vec v^{\thinspace \prime}/c$.

For low emission altitudes ($r\ll\rlc$) and near the dipole axis
(small angle approximation) eq.~\mref{aberration} can be considerably
simplified because $\vec k^\prime$ is not far from radial and
$\vec v_{\rm crt}$ is small and purely azimuthal.
Using the spherical coordinates $(r,\phi,\theta)$, with $\hat z$-axis 
along $\vec \Omega$ and with the azimuth $\phi$
increasing in the direction of the corotation velocity $\vec v_{\rm crt}=\vec \Omega\times \vec r$
the aberration formula becomes:
\begin{eqnarray}
\vec k = (1,\phi_k,\theta_k) &\approx &(1,\phi_{k^\prime}^\prime+r/\rlc,
\theta_{k^\prime}^\prime) = \nonumber \\
&=& \vec k^\prime + (0,r/\rlc,0),
\label{aberration2}
\end{eqnarray}
which means that in our specific case the aberration basically rotates
the emission direction forward in phase by $r/\rlc$ rad, regardless
of the emission colatitude 
(DRH).
Note that eq.~\mref{aberration2} refers to the unit vectors of emission
direction after they have been parallel-shifted to the origin of the 
coordinates' frame: $\vec k = (1,\phi_k,\theta_k)$, $\vec k^\prime =
(1,\phi_{k^\prime}^\prime,\theta_{k^\prime}^\prime)$.

Propagation time delays (also called `retardation') refer to
the travel time of radio waves within the pulsar magnetosphere:
radiation emitted deeper in the magnetosphere will be detected later 
than the simultaneous emission from high altitudes.
The delays are taken into account through 
the reference to the center of the neutron star (NS).
Let us consider a point located at the radial position $\vec r$
at the instant $t_{\rm em}$ when it is emitting towards an observer.
Under normal conditions such a point is located closer to the observer
than the NS center. 
Radiation emitted from this point
will therefore be detected \emph{earlier} by $\vec r(t_{\rm em}) \cdot \nobs/c$
than an imaginary signal simultaneously emitted from the center of the NS.
Here $\vec r(t_{\rm em}) \cdot \nobs$ is a projection of $\vec r$ 
(taken at the instant of emission) on the observer's line of sight 
(defined by the unit vector $\nobs$) and I assume that the radio 
waves travel at the speed of light $c$ (dispersion is ignored). 
Note that the propagation time advance/delay
is \emph{not} independent of aberration: to find $\vec r(t_{\rm em})$
at the moment of emission, we first need to know the aberrated emission 
direction $\vec k(\vec r)$. 
However,
since the angle between $\vec r(t_{\rm em})$ and $\nobs$ is small,
in the lowest order approximation 
the propagation time advance simply becomes: $\vec r(t_{\rm em}) 
\cdot \nobs/c \approx r/c$ and the associated phase advance is $r/\rlc$,
just as in the case of aberration.

\bsp

\label{lastpage}

\end{document}